\documentclass[twocolumn,fleqn,usenatbib]{aastex61}

\usepackage{graphicx} 
\usepackage{epsfig}
\usepackage{float} 
\usepackage{color} 
\usepackage{ulem}
\usepackage{lastpage}   
\usepackage{breakurl}

\usepackage{amsmath}	
\usepackage{amssymb}	
\usepackage{aas_macros}

\def\starlight{\textsc{starlight}}                    

\newcommand {\be} {\begin{equation}}
\newcommand {\ee} {\end{equation}}

\newcommand{\msun}{~M$_{\odot}$}

\newcommand{\Mstar}{$M_{\star}$}

\newcommand{\MBH}{$M_{\rm BH}$}
\newcommand{\Lrad}{$L_{\rm 1.4}$}
\newcommand{\LHa}{$L_{\Ha}$}

\newcommand{\Lbol}{$L_{\rm bol}$}
\newcommand{\Avneb}{$A_{V}^{\rm neb}$}
\newcommand{\Avstar}{$A_{V}^*$}

\newcommand{\hii}{\ifmmode \rm{H}\,\textsc{ii} \else H~{\sc ii}\fi}
\newcommand{\Ha}{\ifmmode {\rm H}\alpha \else H$\alpha$\fi}
\newcommand{\Hb}{\ifmmode {\rm H}\beta \else H$\beta$\fi}
\newcommand{\oiii}{\ifmmode [\rm{O}\,\textsc{iii}] \else [O~{\sc iii}]\fi}
\newcommand{\oii}{\ifmmode [\rm{O}\,\textsc{ii}] \else [O~{\sc ii}]\fi}
\newcommand{\oi}{\ifmmode [\rm{O}\,\textsc{i}] \else [O~{\sc i}]\fi}
\newcommand{\nii}{\ifmmode [\rm{N}\,\textsc{ii}] \else [N~{\sc ii}]\fi}
\newcommand{\sii}{\ifmmode [\rm{S}\,\textsc{ii}] \else [S~{\sc ii}]\fi}
\newcommand{\Oiii}{[O~{\sc iii}]$\lambda$5007}
\newcommand{\Nii}{[N~{\sc ii}]$\lambda$6584}

\shorttitle{The host galaxies of radio-loud and radio-quiet AGNs}
\shortauthors{Kozie\l-Wierzbowska et al.}

\begin{document}

\title{What distinguishes the host galaxies of radio-loud and radio-quiet AGNs?}

\correspondingauthor{D. Kozie\l-Wierzbowska}
\email{dorota.koziel@uj.edu.pl}

\author{D. Kozie\l-Wierzbowska}
\affil{Astronomical Observatory, Jagiellonian University, ul. Orla 171, PL-30244 Krakow, Poland}

\author{N. Vale Asari}
\affil{Departamento de F\'{\i}sica--CFM, Universidade Federal de Santa Catarina, C.P.\ 476, 88040-900, Florian\'opolis, SC, Brazil}

\author{G. Stasi\'nska}, 
\affil{LUTH, Observatoire de Paris, CNRS, Universit\'e Paris Diderot; Place Jules Janssen F-92190 Meudon, France}

\author{M. Sikora},
\affil{Nicolaus Copernicus Astronomical Center, Bartycka 18, 00-716 Warsaw, Poland}

\author{E. I. Goettems}, 
\affil{Departamento de F\'{\i}sica--CFM, Universidade Federal de Santa Catarina, C.P.\ 476, 88040-900, Florian\'opolis, SC, Brazil}
\author{A. W\'ojtowicz}
\affil{Astronomical Observatory, Jagiellonian University, ul. Orla 171, PL-30244 Krakow, Poland}

\begin{abstract}
We compare the optical properties of the host galaxies of radio-quiet (RQ) and
radio-loud (RL) Type 2 active galactic nuclei (AGNs) to infer whether the  jet
production efficiency depends on the host properties or is determined just by intrinsic
properties of the accretion flows. We carefully  select  galaxies from
SDSS, FIRST, and NVSS catalogs.  We confirm  previous findings that
the fraction of  RL AGNs depends on the black-hole (BH) masses, and on the Eddington ratio. 
The comparison of the nature of the hosts of RL
and RQ AGNs, therefore, requires pair-matching techniques. 
Matching in BH mass and Eddington ratio allows us to study the differences between galaxies hosting RL and RQ AGNs
that have the same basic accretion parameters.
We show that 
these two samples differ predominantly in the host-galaxy concentration index,
morphological type (in the RL sample the frequency of elliptical galaxies becoming larger
with increasing radio loudness), and nebular extinction (galaxies with highest radio loudness showing only low nebular extinction).  
Contrary to some previous studies, we find  no significant
difference between our radio-loud and radio-quiet samples regarding
merger/interaction features.
\end{abstract}



\section{Introduction}
\label{sec:Introduction}

\defcitealias{Best.Heckman.2012a}{BH12}
\defcitealias{Kewley.etal.2001a}{K01}

At least 10\% of active galactic nuclei (AGNs) is associated with radio sources powered by jets  \citep[][and references therein]{Kellermann.etal.2016a}. The radio loudness of these AGNs --  defined to be the ratio of the radio flux to the optical flux -- covers three to four orders of magnitude (\citealp[e.g.][]{Sikora.Stawarz.Lasota.2007a, Lal.Ho.2010a}). This implies very diverse jet production efficiencies.  For jets powered by rotating black-holes \citep[BHs;][]{1977MNRAS.179..433B}, such a diversity can result from the spread of the BH spins and magnetic fluxes. If the values of these parameters are mainly determined by the cosmological
evolution of the BH and its environment prior to the AGN/quasar phase, then one might expect to see correlations between the radio loudness and some properties of the host galaxies and their environments. And, indeed, such correlations have been indicated by several independent studies using different samples. It was claimed that the most radio-loud AGNs are preferentially hosted by bulge-dominated galaxies with masses larger than $10^{11} M_{\odot}$ and BH masses larger than $10^8 M_{\odot}$ (\citealp[e.g.][]{Laor2000a, Dunlop.etal.2003a, Floyd.etal.2004a, McLure.Jarvis.2004a, Best.etal.2005a}); the fraction of galaxies with disturbed morphology is larger in RL
AGNs than in RQ AGNs (\citealp{Bessiere.etal.2012a,Chiaberge.etal.2015a}); the star-formation rate (SFR)
 in the hosts of RL AGNs is lower than in the hosts of RQ
 AGNs (\citealp{Dicken.etal.2012a, Floyd.etal.2013a}); the environment is denser around RL AGNs than around RQ AGNs (\citealp{Mandelbaum.etal.2009a, Shen.etal.2009a, Donoso.etal.2010a, Falder.etal.2010a, RamosAlmeida.etal.2013a}).
The dependence of radio loudness on host-galaxy properties is also indicated by studies based on samples selected from massive optical and radio surveys \citep{Kauffmann.Heckman.Best.2008a, Best.Heckman.2012a, Gurkan.etal.2015a}. In the latter
 studies, the comparison of the host properties were performed by pairing radio-loud AGNs with radio-quiet AGNs in redshift, stellar mass, and velocity dispersion ($\sigma_{\star}$). Because, for such a set of parameters, the results can be significantly biased by the fact that the paired objects may have very different Eddington ratios, we decided to perform similar studies but pairing radio-loud and radio-quiet AGNs in redshift, BH mass (\MBH, given by $\sigma_{\star}$), and the Eddington ratio $\lambda$. 

In this paper, we concentrate on Type 2 (i.e. obscured
  AGNs) with Eddington ratios $\lambda > 0.003$ and we look for
  differences between galaxies hosting AGNs with radio emission
  associated with a jet activity, and galaxies not detected in radio
 (see Section \ref{sec:radio}). 
By limiting ourselves to obscured AGNs, we avoid
pollution of the spectra by the broad H$\beta$ and Fe~{\sc ii} lines
arising in the vicinity of BHs as well as contamination of the stellar
continuum by the emission from the AGNs. And by limiting ourselves to
$\lambda > 0.003$, we avoid in our sample AGNs
with radiatively inefficient accretion flows  (\citealp[e.g.][]{Best.Heckman.2012a, Stern.Laor.2013a}). 

Our master sample of galaxies is the seventh release of
the Sloan Digital Sky Survey (SDSS DR7, \citealp{Abazajian.etal.2009a}).
Our master sample of radio galaxies is the
\citet[][BH12]{Best.Heckman.2012a} catalog, obtained by cross-matching
the DR7 Main Galaxy Sample \citep{Strauss.etal.2002a} and Luminous Red
Galaxy Sample \citep{Eisenstein.etal.2001a} with radio sources from the
NRAO VLA Sky Survey (NVSS, \citealp{Condon.etal.1998}) and the Faint
Images of the Radio Sky at Twenty-centimeter (FIRST,
\citealp{Becker.etal.1994a}) catalogs\footnote{Data considered in this paper are available as ascii files at \href{https://doi.org/10.5281/zenodo.835591}{10.5281/zenodo.835591}}.

The organization of the paper is as follows. In Section \ref{sec:data},
we explain how the final database was selected and describe how 
the
parameters necessary for our analysis were obtained. In Section
\ref{sec:character}, we characterize our samples of radio-loud and
radio-quiet galaxies. In Section \ref{sec:match}, we compare the
photometric and spectroscopic properties of the two samples by using a
pairing technique. In Section \ref{sec:morph}, we compare the
morphological properties of the galaxies in our radio-loud sample and in
the matched galaxies of our radio-quiet sample. In Section
\ref{sec:conclus}, we summarize our results and speculate on possible
interpretations.

Throughout the paper, we consider a $\lambda$CDM cosmology with
$H_0 = 70 \ {\rm km} \ {\rm s}^{-1} \ {\rm Mpc}^{-1}$, $\Omega_m=0.30$,
and $\Omega_{\lambda}=0.70$.
 
 
\section{The data}
\label{sec:data}

\subsection{Optical}
\label{sec:optical}

The galaxies are selected from the SDSS DR7 database \citep{Abazajian.etal.2009a} with the criteria described in Sect.  \ref{sec:opticalsamples}.  The SDSS DR7 spectrophotometric calibration is inadequate for extended sources, so we applied a flux renormalization correction to match the spectral flux to the fiber photometry in the $r$ band (the `spectofibre' factor provided by the MPA-JHU team at \url{http://wwwmpa.mpa-garching.mpg.de/SDSS/DR7/raw_data.html}, as in, e.g. \citealp{Thomas.etal.2013a}).  The SDSS spectra have been processed with the spectral synthesis code \starlight\ as described below.

\subsubsection{Data processing with \starlight\ and derived quantities}
\label{sec:starlight}

\starlight\ \citep{CidFernandes.etal.2005a} is an inverse stellar
population synthesis code that recovers the stellar population of a
galaxy by fitting a pixel-by-pixel model to the observed spectrum
(excluding bad pixels, narrow windows where emission lines are expected,
and the region of the Na D doublet). The model is a linear combination
of 150 simple stellar populations' templates of given age $t_\star$ and
metallicity $Z/Z_\odot$. The ages range between 1 Myr and 18 Gyr, and
the metallicities between 0.005 and 2.5. The templates are obtained in
the same way as in \cite{CidFernandes.etal.2010a}, i.e.  using
\citet{Bruzual.Charlot.2003a} evolutionary stellar population models,
with the STELIB library of stellar atmospheres
\citep{LeBorgne.etal.2003a}, `Padova 1994' stellar evolution tracks
\citep{Bertelli.etal.1994a}, and \citet{Chabrier.2003a} initial mass
function.

The stellar dust attenuation \Avstar\ is obtained by \starlight,
adopting a \citet*{Cardelli.Clayton.Mathis.1989a} extinction law with
$R_V = 3.1$ by requiring that the reddened modeled spectrum matches the
observed one.

The intensities of the emission lines were measured by Gaussian fitting
after subtracting the modeled stellar spectrum from the observed one,
which eliminates contamination by stellar features.

The nebular extinction \Avneb\ was computed from the measured \Ha/\Hb\
emission-line ratio by assuming an intrinsic ratio of $3$ 
and a
\citet{Cardelli.Clayton.Mathis.1989a} law with $R_V = 3.1$ and the
emission-line fluxes were then corrected for nebular extinction. More
details on the adopted procedures can be found in
\citet{Mateus.etal.2006a}, \citet{Stasinska.etal.2006a},
\citet{Asari.etal.2007a} and \citet{CidFernandes.etal.2010a}.

All the data used in this paper can be retrieved from the \starlight\
database\footnote{http://www.starlight.ufsc.br}
\citep{CidFernandes.etal.2009a}.

The total stellar masses of the galaxies, \Mstar, were obtained as in
\citet{CidFernandes.etal.2005a}, assuming that the mass-to-light ratios
are the same outside and inside the fiber and scaling the stellar masses
encompassed by the fiber by the ratio between total (from the
photometric database) and fiber $z$-band luminosities. This correction
is smaller than a factor of two in a large portion of our sample, but can
amount to factors of up to eight.  On the other hand, we do not correct the
emission-line luminosities for aperture effects, since the emission
lines are expected to be emitted mainly in the inner regions of the
considered galaxies (see however the proviso expressed in Section
\ref{sec:opticalsamples}).  We also make use of some parameters related
to the star-formation histories and stellar mass growth extracted from
the \starlight\ database and explained in the next sections.

To convert \LHa\ to \Lbol\ we follow \citet[equation 1]{Netzer.2009a}, 
using the expression
\begin{equation}
\label{eq: Lbol}
\log (L_{\rm bol} / L_{\odot}) = \log (L_{\Ha}/L_{\odot}) + 3.01 + C,
\end{equation}
where $C \equiv \text{max}\{ 0.0,\, 0.31 (\log \oiii/\Hb - 0.6) \}$.

The black-hole mass, \MBH, is estimated from the stellar
velocity dispersion determined by \starlight, $\sigma_{*}$,
using the relation by \citet{Tremaine.etal.2002a}:
\begin{equation}
\label{eq: MBH}
\log (M_\mathrm{BH} / M_{\odot}) = 8.13 + 4.02 \log(\sigma_\star / 200 \mathrm{\;km\;s}^{-1}).
\end{equation}

The Eddington ratio is defined as $\lambda = L_{\rm bol}/L_{\rm Edd}$,
where {$L_{\rm Edd} \,[{\rm L_\odot}] = 3.3 \times 10^4 M_{\rm BH} \,[{\rm M_\odot}]$
\citep[equation 4]{Heckman.Best.2014a}.

\subsubsection{The optical samples}
\label{sec:opticalsamples}

\begin{enumerate}

\item We consider galaxies belonging either to the Main Galaxy Sample  \citep{Strauss.etal.2002a} or to the Luminous Red Galaxy sample \citep{Eisenstein.etal.2001a}.

\item In order to allow a reliable analysis of the stellar populations, we require a  signal-to-noise ratio in the continuum at 4020 \AA\   of at least 10 (see justification in \citealp{CidFernandes.etal.2005a}).

\item A small fraction of the remaining galaxies (0.2\%) for which the Petrosian half-light radius is negative or the stellar mass is smaller than $10^7 \mathrm{M}_\odot$ in the \starlight\ database are also eliminated.

\end{enumerate}

These criteria select 673,807 galaxies. 

Further criteria are adopted to extract galaxies containing an AGN and for which the treatment we apply (in particular, the estimation of the Eddington ratio) are expected to be valid.

\begin{enumerate}

\item We impose the redshift $z$ to be larger than 0.002 to guarantee that luminosity distances are not dominated by peculiar motions \citep[e.g.][]{Ekholm.etal.2001a} and smaller than 0.4  to include the \Ha\ line in the spectrum.

\item We remove galaxies with stellar velocity dispersions smaller than 70 km s$^{-1}$ to obtain meaningful black-hole masses.

\item We remove galaxies with faulty pixels around important diagnostic emission lines, requiring at least 75\% of good pixels at one sigma from the peak of \Oiii, \Nii, \Ha, and \Hb\ (the same criterion as used in \citealp{Stasinska.etal.2015a}).

\item We impose an S/N in those lines of at least 1.5. With such a low limit on the S/N, we cannot ensure that the position in the \oiii/\Hb\ versus \nii/\Ha\ diagram (the so-called BPT diagram, after \citealp{Baldwin.Phillips.Terlevich.1981a}) is  always accurate, but it is sufficient for our needs. The presence of \Hb\ is necessary  to correct the line emission for extinction.

\item We keep only galaxies that lie above the \citetalias{Kewley.etal.2001a} line (after \citealp{Kewley.etal.2001a}) in the BPT diagram to remove galaxies dominated by star formation.

\item We remove galaxies that could
 be `retired' galaxies according to the EW(\Ha) versus \nii/\Ha\ diagram (the WHAN diagram, see \citealp{CidFernandes.etal.2011a}). Although these galaxies lie in the zone of LINERs in the BPT diagram, their emission lines can  be
 produced by hot low-mass evolved stars (HOLMES) and not by an AGN. Since we have no way to distinguish a priori in which galaxies the \Ha\ emission is due to gas ionized by HOLMES and in which ones it is due to a weak AGN, we consider it safer, for the purpose of this paper,  to remove all the galaxies with EW(\Ha) $< 3$.
 Note that this criterion drastically reduces the sample from 76,077 to 19,883 galaxies.

\end{enumerate}
  
This is our main sample and comprises 19,883 objects. 

\begin{table*}
\caption{The different samples and subsamples considered in this paper.}
\label{tab:sample}
\begin{center}
\begin{footnotesize}
\begin{tabular}{lllrl}
\hline
Sample        & Description                          & Color in figs    & N         & Criteria \\
\hline                                          
main          & --                                   & --                & $19,883$ & SDSS MGS or LRG, $S/N_c \ge 10$, $S/N_\mathrm{BPT} \ge 1.5$, \\ 
                                                                                  &&&& no sky contamination, above \citetalias{Kewley.etal.2001a}, $W_{H\alpha} > 3$ \AA, \\ 
                                                                                  &&&& $0.002 \le z \le 0.4$, $\sigma_\star \ge 70$ km s$^{-1}$ \\
\Lrad         & Radio-detected                       & --                & $ 1,101$ & subset of main, in the \citetalias{Best.Heckman.2012a} catalog \\ &&&& (detected in \Lrad\ by FIRST or NVSS) \\
\Lrad\ AGN    & Radio AGN                            & green             & $    617$ & subset of \Lrad\, classified as AGN by \citetalias{Best.Heckman.2012a}\\
RL            & Radio-loud AGN                       & red               & $    376$ & subset of \Lrad\ AGN, misidentifications removed, $\lambda \ge 0.003$ \\
RQ            & Radio-quiet AGN                      & light blue        & $10,918$ & subset of main, not in the \citetalias{Best.Heckman.2012a}, $\lambda \ge 0.003$, ${\cal R}^{(upp)}<15.8$ \\
mRQ           & Pair-matched radio-quiet AGN         & dark blue         & $    911$ & subset of RQ, pair-matched to RL \\
CRL           & Cleaned radio-loud AGN               & red               & $    134$ & subset of RL, cleaned from SF contamination \\
CRQ           & Cleaned radio-quiet AGN              & light blue        & $ 3,013$ & subset of RQ, cleaned from SF contamination \\
mCRQ          & Pair-matched cleaned radio-quiet AGN & dark blue         & $    296$ & subset of CRQ, pair-matched to CRL \\
\hline
\end{tabular}
\end{footnotesize}
\end{center}
\end{table*}

\subsection{Radio}
\label{sec:radio}

\subsubsection{Catalogs}
\label{sec:radiodata}

We consider the sample of 18,286 radio galaxies from \citet{Best.Heckman.2012a} downloaded from \url{http://vizier.cfa.harvard.edu/viz-bin/VizieR?-source=J/MNRAS/421/1569}. This sample was obtained by refined automatic procedures to cross-match the SDSS DR7 galaxy sample with radio sources from the NVSS and FIRST catalogs. Note that the SDSS catalogs used by \citeauthor{Best.Heckman.2012a} are the same as the ones in our project (Sec. \ref{sec:opticalsamples}). Note also that the FIRST catalog was designed to cover the same sky area as SDSS.
 The cross-matching goes down to a flux density of 5 mJy, corresponding to radio luminosities of \Lrad $= 1.27 \times 10^{23}$ W Hz$^{-1}$ at redshift $z = 0.1$.  In their work, \citeauthor{Best.Heckman.2012a} classified galaxies into star forming (SF) or AGN depending on whether the radio emission is produced by AGN jets or is identified with star-formation processes\footnote{In the \citet{Best.Heckman.2012a} catalog AGNs were separated from star-forming galaxies using a combination of criteria involving the relationship between the 4000 \AA\ break strength and the ratio of radio luminosity per stellar mass, the position in the BPT diagram, and the relation between the \Ha\ emission-line luminosity and the radio luminosity. The total number of radio-AGNs in the \citeauthor{Best.Heckman.2012a} catalog is 15,300 out of 18,286 objects.}.  
We have removed from the present study those radio sources classified by \citet{Best.Heckman.2012a} as star forming and kept only those objects that were classified by them as AGNs. They form what we call the \Lrad\ AGN sample.

We have analyzed the radio and optical images of these objects and found a few cases where the SDSS galaxy associated by \citeauthor{Best.Heckman.2012a} to a radio source is a misidentification. These are SDSS\,J150134.73+544734.0, SDSS\,J123959.04+370505.1, SDSS\,J154322.93+225036.0 and SDSS\,J111209.78+194052.5. Those objects were removed from further consideration. On the other hand, we noted that the \citet{Best.Heckman.2012a} catalog lacks 78 of the extended radio sources presented in \citet{Sikora.etal.2013a}.  Most of these missing radio galaxies are genuine FR II or FR I radio galaxies \citep{Fanaroff.Riley.1974a} with large radio fluxes
(the starting catalogs for the cross-matching of radio with optical data in \citealp{Sikora.etal.2013a} were the Cambridge catalogs, which are significantly shallower than FIRST or NVSS). This suggests that the \citet{Best.Heckman.2012a} catalog might also be missing many extended radio sources with weak total radio fluxes. It is difficult to assess, at the present stage, how the results of this paper would be changed if one had a more complete catalog of radio sources.  We therefore consider our results still preliminary, until a more complete catalog of radio galaxies is produced.

\subsubsection{Samples}
\label{sec:rlsample}
 
For the remainder of the study, we only consider objects with $\lambda \ge 0.003$. Our radio-loud sample (from now on referred to as the RL sample) is composed of all the galaxies from the \Lrad\ AGN sample that have $\lambda \ge 0.003$ and contains 376 objects. Our radio-quiet sample (from now on referred to as the RQ sample) is composed of all the objects in our main sample that have $\lambda \ge 0.003$, which are not in the \citetalias{Best.Heckman.2012a} catalog, 
and whose radio loudness parameter (defined by  \citet{Sikora.etal.2013a} to be ${\cal R} \equiv L_{\rm 1.4} [W\, Hz^{-1}] / L_{\Ha} [ L_{\odot}]$, where $L_{\rm 1.4}$ is the radio luminosity at 1.4 GHz) estimated from the detection limit of the radio catalogs is such that log \ ${\cal R} < 15.8$.\footnote{ The radio loudness defined by \citet{Kellermann.etal.1989a} is ${\cal R}^{(K)} = L_{\rm 5} / L_{\nu_B}$, where $L_{\rm 5}$ is the radio luminosity at 5 GHz. For a radio spectral index $\alpha_R = 0.8$, using equation \ref{eq: Lbol} and $L_{\rm bol} = 5 \times (\nu_B L_{\nu_B})$ \citep{2012MNRAS.422..478R}, we obtain ${\cal R}^{(K)} \simeq 1.6 \times 10^{-15} {\cal R} $. This implies that the commonly used criterion to coin `radio-loud' AGNs, ${\cal R}^{(K)} > 10$, translates into $\log {\cal R} > 15.8$. 
}
It has 10,918 objects.  
 For the reader's convenience Table \ref{tab:sample} summarizes the various samples considered in this paper.

\begin{figure}
\centering
\includegraphics[scale=0.65]{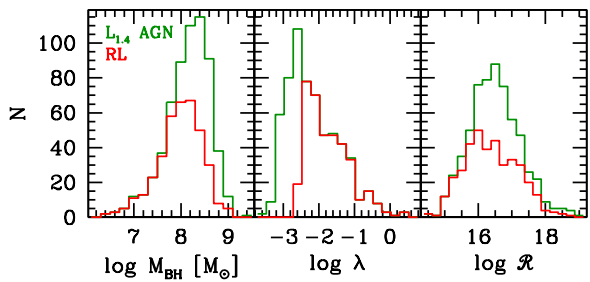}
\vspace{-9cm}
\caption{Histograms with respect to the black-hole mass, Eddington ratio and radio loudness parameter ${\cal R}$. The radio sources from
  \citet{Best.Heckman.2012a} classified as AGNs (the \Lrad\ AGN sample) are in green and the RL sample is in red.}
\label{fig:L1.4RL}
\end{figure}

Our notations RL and RQ do not correspond exactly to the usual definition of radio-loud and radio-quiet. We do use Kellermann's criterion of radio loudness to eliminate from the RQ sample those objects that may have radio jets but fall below the detection limit of FIRST and NVSS surveys. But, on the other hand, we include in our RL sample jetted AGNs that have ${\cal R}$ lower than Kellermann's limit. Therefore, in the RL sample, we have radio-detected AGNs that are not radio-loud according to the usual definition while in the RQ sample we have only intrinsically radio-quiet objects.

Figure \ref{fig:L1.4RL} shows the histograms of the \Lrad\ AGN and RL samples as a function of the BH mass (left panel), Eddington ratio (middle panel), and radio loudness ${\cal R}$ (right panel). 
It can be seen that limiting the RL sample by $\lambda \ge 0.003$, i.e. keeping mostly sources with radiatively efficient accretion, we exclude many
 radio sources with high BH masses and high values of ${\cal R}$.


\section{Characterization of Our RL and RQ samples}
\label{sec:character}

Before proceeding to the pair-matching, it is interesting to investigate the global properties of our
 RL and RQ samples.

\begin{figure}
\centering
\includegraphics[scale=0.65]{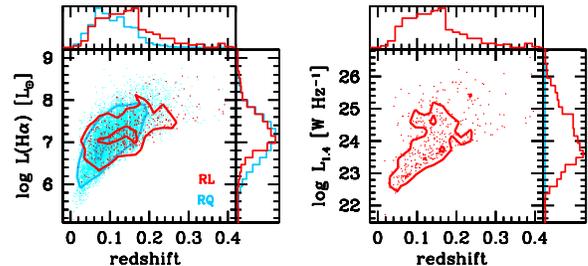}
\vspace{-0.5cm}
\caption{Left panel: the observed \Ha\ luminosity versus redshift for RL and RQ objects (red and blue points, respectively). Right panel: the radio luminosity, \Lrad\, versus redshift for RL objects (red points). The contours in these diagrams correspond to 20\% and 80\% of the objects.  The normalized histograms shown on both axes use the same colors.}
\label{fig:LHaL1.4z}
\end{figure}

The left  panel of Fig. \ref{fig:LHaL1.4z} shows the observed (i.e.\ not corrected for extinction)  \Ha\ luminosity as a function of redshift for the RL and RQ objects (represented in red and blue, respectively).  The contours in these diagrams correspond to 20 and 80\% of the objects. The normalized histograms shown on both axes use the same colors.   The increase with redshift of the lower envelopes of the samples is due to the limitation in magnitude of the SDSS galaxies. The fact that the \Ha\ luminosities of the RL samples do not reach the lowest values of the RQ sample is due to the  \cite{Best.Heckman.2012a} catalog being limited by a radio flux of 5 mJy.  The right panel of Figure \ref{fig:LHaL1.4z} shows the radio luminosity, \Lrad , of the RL objects as a function of redshift. Again the increase of the lower envelope with redshift is  due to the flux limit of the \cite{Best.Heckman.2012a} catalog.

\begin{figure*}
\centering
\includegraphics[scale=0.9]{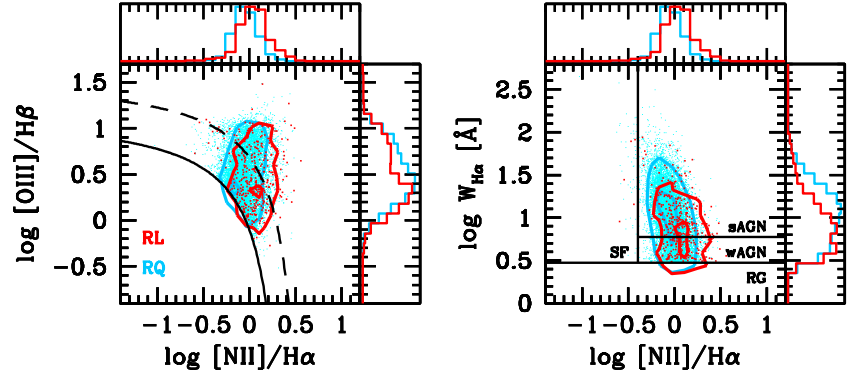}
\caption{BPT and  WHAN diagrams for our  RL and RQ objects, using the same representation as in Fig. \ref{fig:LHaL1.4z}. In the BPT diagram, the continuous black curve is the \citetalias{Kewley.etal.2001a} line. The dashed black curve corresponds to equation \ref{eq: shifted} and delimits our cleaned samples, CRL and CRQ.}
 \label{fig:BPT}
\end{figure*}

Figure \ref{fig:BPT} shows the positions of the RL and RQ objects in the BPT diagram (left) and in the WHAN diagram (right),  using the same layout as   Fig. \ref{fig:LHaL1.4z} (left). We can see that the distribution of the points in the BPT and WHAN diagrams is very similar for both samples, suggesting that the ionization conditions of the emitting gas are very similar. 

\begin{figure*}
\centering
\includegraphics[scale=0.9]{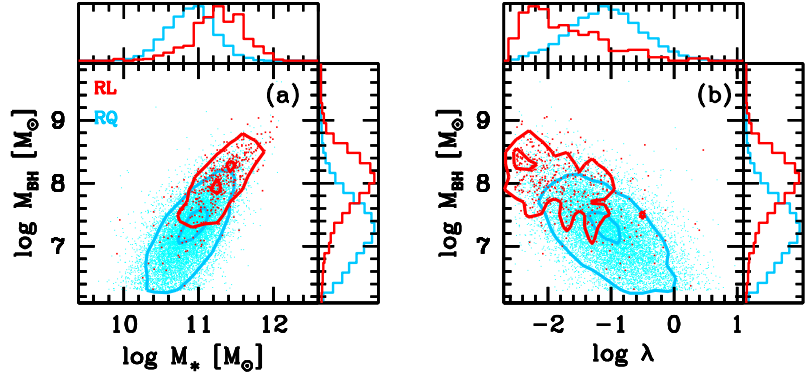}
\caption{Distribution of our RL and RQ samples in several 2D
  diagrams: BH mass (\MBH) versus stellar mass ($M_\star$), and \MBH\ versus Eddington
  ratio, ($\lambda = L_{\rm bol}/L_{\rm Edd}$). The figure uses the same layout as Fig. \ref{fig:LHaL1.4z}.}
 \label{fig:char-1}
\end{figure*}

Figure \ref{fig:char-1} shows plots characterizing our RL and RQ samples, with the same layout as Fig. \ref{fig:LHaL1.4z} (left). The left panel shows the values of \MBH\ versus the galaxy stellar mass, \Mstar, the right panel shows those of \MBH\ versus $\lambda$.  We see that \MBH\ and \Mstar\ are strongly correlated in our samples, as found in numerous studies related to classical bulges and ellipticals \citep[e.g.][]{Ferrarese.Merritt.2000a, Best.etal.2005a, Gadotti.Kauffmann.2009a, Kormendy.Ho.2013a}, and that RL objects have on average higher values of \MBH\ and \Mstar. This implies
 that, to compare the properties of RL and RQ galaxies, pairing in either \MBH\ or \Mstar\ is needed, but that it is not necessary to pair both in \MBH\ and \Mstar.  The right panel illustrates the trends mentioned in the Introduction that there is a much larger fraction of radio-loud AGNs at larger BH masses and lower Eddington ratios.  Sources with small $\lambda$ and small \MBH\ are not visible there because these are objects with \Ha\ fluxes too faint to be detected in SDSS spectra.  On the other hand, sources with high $\lambda$ and high \MBH\ are rare. We see that at a given value of \MBH\ AGNs can have values of $\lambda$ in a range of two orders of magnitude, implying that it is necessary to pair the galaxies also in $\lambda$.

\begin{figure*}
\centering
\includegraphics[scale=0.9]{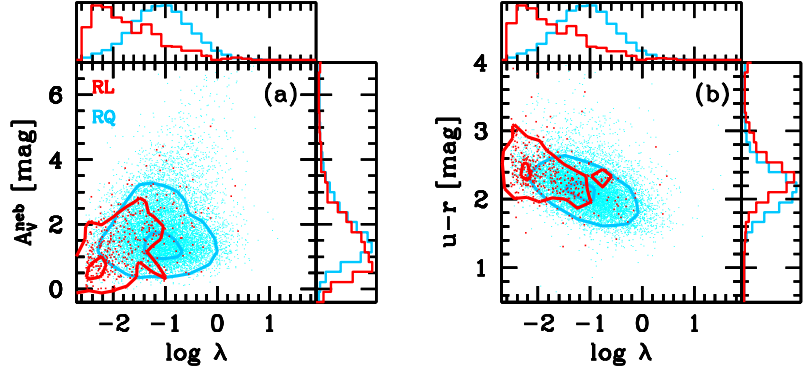}
\caption{Nebular extinction $A_{V}^{\rm neb}$ and the color $u-r$ versus $\lambda$ for our RL and RQ samples.  The figure uses the same layout as Fig. \ref{fig:LHaL1.4z}.}
\label{fig:char-2}
\end{figure*}

Figure \ref{fig:char-2} shows the values of the extinction derived from \Ha\ and \Hb\ ($A_{V}^{\rm neb}$, panel a) and of the $u-r$ color (panel b) as a function of $\lambda$, with the same layout as Fig. \ref{fig:LHaL1.4z}. The histograms of $A_{V}^{\rm neb}$ and $u-r$ suggest that the RL and RQ samples do not differ in $A_{V}^{\rm neb}$ but differ somewhat in $u-r$.  However, as will be shown in the next section, one needs to be careful with the interpretation of such histograms since they only describe the properties of global samples.


\section{Comparison of the properties of matched RL and RQ samples}
\label{sec:match}

If we wish to investigate what distinguishes RL and RQ AGNs, we need, as argued before, to apply a matching technique such as used by \citet{Kauffmann.Heckman.Best.2008a, Masters.etal.2010a, Wild.Heckman.Charlot.2010a}, but with carefully selected parameters to be matched. To each object of the RL sample, we associate one or several objects from the RQ sample that have very similar values of \MBH, $\lambda$ and $z$, i.e. similar AGN properties, and we look for systematic differences in the remaining properties of those objects between the RL and RQ samples. The matching in $z$ is necessary to ensure that the covering fraction of the SDSS spectroscopic fiber is similar in any pair of galaxies from the RL and RQ sample. It also ensures that the morphological classification of RL and matched RQ galaxies will be equally biased. The matching in $\lambda$ ensures that we compare objects with similar rates of activity.

In practice, the matching is done as follows.  For each RL object, we select all the RQ objects that satisfy $|\Delta z| \le 0.03$, $|\Delta \log \lambda| \le 0.15$, and $|\Delta \log M_\mathrm{BH}| \le 0.15$. This procedure selects, on average, 61 suitable pairs for each RL parent object. 62 galaxies in the RL sample have no counterpart, 12 have only one match, 7 only two matches, and the remaining 295 parents have at least 3 suitable pairs. We compute the distances of these RQ objects to their parent RL one according to {$d^2 = (\Delta z /r_z)^2 + (\Delta \log \lambda /r_\lambda)^2 + (\Delta \log M_\mathrm{BH} /r_ {M_{\mathrm{BH}}})^2$},
where  $r_z$, $r_\lambda$ and $r_ {M_{\mathrm{BH}}}$ are the observed ranges of the parameters $z$, $\log\ \lambda$ and $\log\ M_\mathrm{BH}$. We then look for the $n$ objects with the lowest values of the distance $d$.

In the following, we show results for $n=3$, which we found to be the optimal compromise to ensure that the statistics are reasonable without reducing the quality of the match. Our matched sample thus contains roughly three times more objects than the parent RL sample.

We note that one single RQ object may be matched to more than one parent RL galaxy. We find 575 objects with only one parent galaxy, 84 objects matched to 2 parents, 22 to 3, 9 to 4, 6 to 5, 6 to 6, and no galaxies matched to 7 parents or more.  We have tested our procedure to a randomly selected subset of the RQ sample to be matched to the remaining RQ. We have verified that the distribution of parameters in the matched RQ sample is very similar to the the original RQ sample, from which we conclude that our pair-matching procedure does not artificially skew our results.

\begin{figure}
\centering
\includegraphics[width=.91\columnwidth]{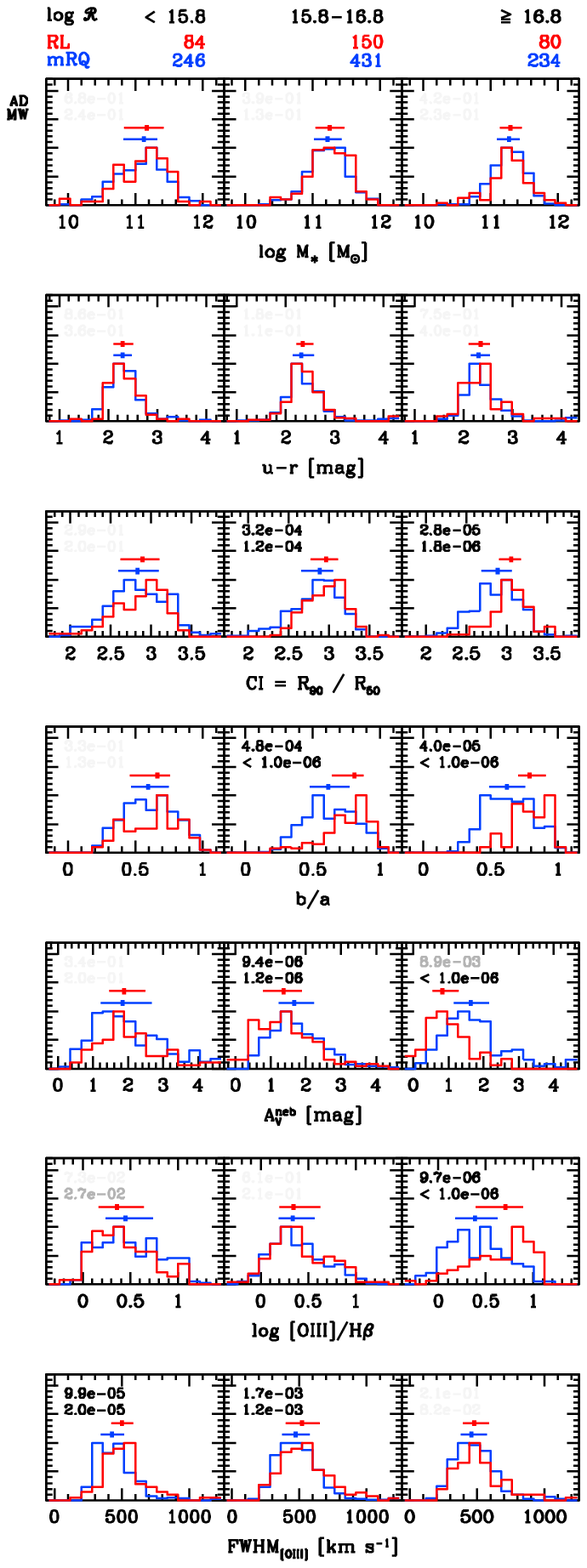}
\caption{Histograms of selected parameters for the objects in the RL sample (in red) and the matched RQ ones (in blue) in three ${\cal R}$ bins. The median values in each sample are marked on top of the histograms; horizontal lines show the quartiles. The range of ${\cal R}$ values is shown at the top, and the number of objects is indicated above each bin in red for RL and in blue for mRQ objects. The numbers in each panel show the results of Anderson-Darling (top value) and Mann-Whitney (bottom value) tests. Results more significant than $p < 0.003$ and  are marked in black, more significant than $p < 0.05$ in dark gray, and others in light gray.  }
 \label{fig:compmatch}
\end{figure}

\begin{figure}
\centering
\includegraphics[width=.91\columnwidth]{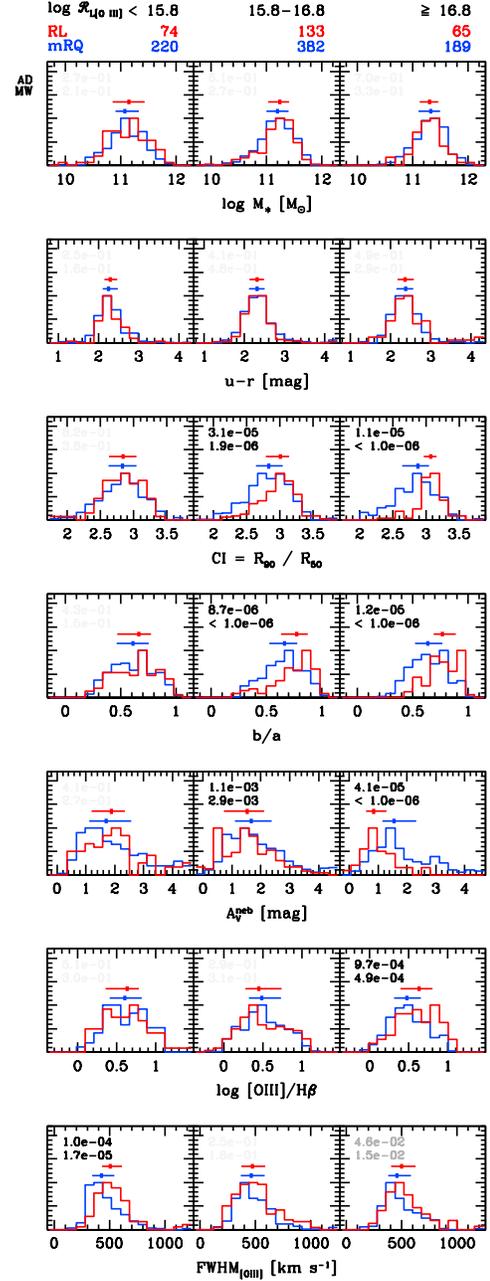}
\caption{Histograms of selected parameters for the objects in the RL sample (in red) in three ${\cal R_{\oiii}}$ bins. Histograms of  the RQ AGNs matched to RL ones from each bin are printed in blue. The layout is the same as in Fig. \ref{fig:compmatch}. Results more significant than $p < 0.003$ and  are marked in black, more significant than $p < 0.05$ in dark gray, and others in light gray.} 
 \label{fig:compmatchdiff}
\end{figure}

\begin{figure}
\centering
\includegraphics[width=.91\columnwidth]{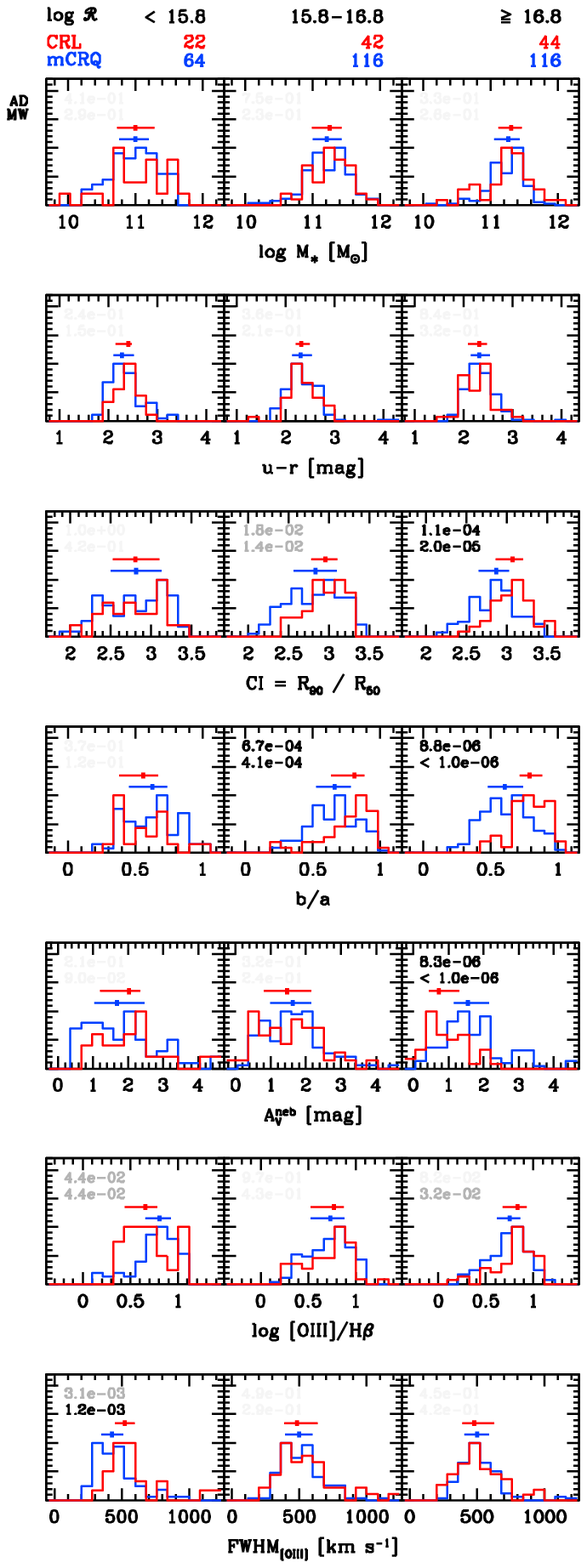}
\caption{The same as Fig.  \ref{fig:compmatch} but for the
 cleaned samples, CRL and mCRQ. } 
 \label{fig:compmatchC}
\end{figure}

Figure \ref{fig:compmatch} shows the histograms of selected parameters for the objects in the RL sample (in red) and matched RQ sample (in blue)  dividing the objects into three bins of the ${\cal R}$ values for the  RL objects
(to see if there are any trends with radio loudness). 
The chosen radio loudness bins are $\log {\cal R} < 15.8$;
$15.8$--$16.8$; and $\log {\cal R} > 16.8$. Such a division is
motivated by the fact that the value ${\log \cal R} = 15.8$
(corresponding to ${\cal R}^{(K)} = 10$) is commonly used to divide
AGNs into RL and RQ samples (see footnote 4), while radio sources whose
radio loudness is ten times higher often have FR II morphologies \citep[see, e.g.][]{Lu.etal.2007a}.  Since the radio emission of
AGNs with $R^{(K)} < 10$ is expected to be associated with star
formation \citep[e.g.][]{Kimball.etal.2011a} or with regions
of energy dissipation of accretion disk winds \citep{Blundell.Kuncic.2007a,Zakamska.etal.2016a}, and because we have
already removed those objects that \citetalias{Best.Heckman.2012a} associated with star
formation (see Sect \ref{sec:radiodata}), the radio emission of our RL AGNs with
${\log \cal R} < 15.8$ is presumably dominated by the winds. In each
radio loudness bin, the median values are materialized by the thick marks on the top of the histograms (red for the RL sample and blue for the matched RQ sample), and the quartiles by the horizontal lines.
 
In each bin, the comparison between RL galaxies and their matched RQ AGNs is made using two statistical tests. The first is the Anderson-Darling (AD) test, and the second is the Mann-Whitney (MW) test. Both give the probability that two samples are drawn from the same population. The values of these probabilities are reported in each panel (top values are for the the AD test).  In black, we mark probabilities more significant than $p<0.003$, and in dark grey $p<0.05$.
 
Of the parameters that were compared in the RL and mRQ samples, only  \Mstar\ and $u-r$ have distributions that are statistically indistinguishable
 ($p > 0.05$ in the AD and MW tests).
The result for the stellar mass is of course expected since we performed the match in $M_\mathrm{BH}$ and since, as seen in Fig.  \ref{fig:char-1}, in our samples \Mstar\ is strongly correlated with $M_\mathrm{BH}$. We note that \Mstar\ increases with ${\cal R}$. We also see that the color $u-r$ does not show any specific trend with ${\cal R}$ and is similar in the two matched samples. This seems contrary to what is seen in Fig. \ref{fig:char-2}. When we consider RL and RQ objects globally, the difference in colors of their host galaxies is noticeable, but when we match RL and RQ sources in AGN properties this difference disappears. 

From Fig. \ref{fig:compmatch} we also see that the values of CI tend to be larger in the RL sample than in the RQ sample in the two bins with higher ${\cal R}$. There is, however, a large overlap. The significance of the difference in the CI distributions is very high ($p < 0.0001$ according to
 the AD and MW tests).

A similar behavior is seen in the case of the host-galaxy axes ratio, $b/a$. RL galaxies have, on average, larger $b/a$ in the two highest ${\cal R}$ bins.  Both CI and $b/a$ refer to the galaxy morphology, lower values of CI and $b/a$ pointing towards more disky galaxies. This is thus a difference
 between hosts of RL and RQ galaxies and should bear some information on the radio loudness phenomenon. We also note that the full-widths at half-maximum of the \oiii\ line are marginally larger for the RL sample than for the mRQ sample, being the largest for the lowest ${\cal R}$ bin. 
 The nebular extinction $A_{V}^{\rm neb}$ decreases from the highest ${\cal R}$ to the lowest bin, and its distribution in the RL and matched RQ samples differs ($p < 10^{-6}$ in the MW test) in the bins of largest ${\cal R}$. Finally, we note that the values of \oiii/\Hb\ are larger in the RL sample than in the mRQ sample in the bin of highest ${\cal R}$ bin (although the difference is only moderately significant).

However, the values of \oiii/\Hb\ and of the luminosity of \Ha\ itself, which we use to estimate the AGN bolometric luminosity, can be affected by \hii\ regions in the host galaxies. One solution is to use the \oiii\ line to estimate the bolometric luminosity.  As the photoionization models of Stasi\'nska et al. (2006) show, this line is less affected by star formation than \Ha\, and, although alone it is not a good proxy of the AGN bolometric luminosity \citep[see][]{Netzer.2009a} because of metallicity and ionization effects, it is widely used by many groups to calculate $L_{\rm bol}$\footnote{The reason for using \oiii\ to compute bolometric luminosities in the literature is actually not that this line is less affected by star formation, but simply that it is the strongest line in the spectra of such objects. As argued by Koziel-Wierzbowska \& Stasi\'nska (2011), this is not a good reason.}.  We therefore constructed Fig. \ref{fig:compmatchdiff}, which is identical to Fig.  \ref{fig:compmatch} but with the bolometric luminosity, Eddington ratio, and radio loudness parameter calculated from the \oiii\ line (to distinguish them from parameters calculated using \Ha, we add the index \oiii).  As can be seen by comparing Fig. \ref{fig:compmatch} and \ref{fig:compmatchdiff} the main results shown in Fig \ref{fig:compmatch} still hold.  As before, the mass distributions of RL and RQ galaxies cannot by distinguished, 
and there is still a significant difference between the RL and mRQ samples in CI and $b/a$.  However, the \oiii/\Hb\ ratio is still different between RL AGNs with $\log {\cal R} > 17$ and the matched RQ galaxies, suggesting that also \oiii\ is also affected by \hii\ regions.  Therefore, we go back to using \Ha\ to measure $L_\mathrm{bol}$ throughout the rest of our paper, and to test how much our results may be affected by the \hii\ regions, we use `cleaned' samples as defined below.

We have to note that the \citetalias{Kewley.etal.2001a} line, which, since \citet{Kauffmann.etal.2003c} is generally considered as delimiting pure AGNs from composite AGNs (i.e. objects where part of the line emission is due to star formation and not to the AGN), may actually correspond to as much as $70\%$ of \Ha\ arising from \hii\ regions \citep{Stasinska.etal.2006a}. If we want AGN hosts that are not -- or very little -- contaminated by star formation, we must select them in the upper-right part of the BPT diagram. In practice, we use a line that is shifted with respect to the \citeauthor{Kewley.etal.2001a} line and whose equation is
\begin{equation}
\label{eq: shifted}
\log \oiii/\Hb = \frac{0.61}{\log \nii/\Ha - 0.67} + 1.59.
\end{equation}
We then define what we call `cleaned' samples, CRL and CRQ, which contain, respectively, 134 and 3\,013 objects.

We now construct Fig.  \ref{fig:compmatchC}, which is identical to Fig.  \ref{fig:compmatch} but with our cleaned samples CRL and mCRQ. The number of sources has decreased by about a factor of two to four compared to Fig.  \ref{fig:compmatch} but it is still high enough to obtain significant results.  We see in Fig.  \ref{fig:compmatchC} that  now \oiii/\Hb\   does not show any difference between the RL and mRQ samples in any ${\cal R}$ bin. The difference in \oiii/\Hb\  in Figure \ref{fig:compmatch} was then probably due to the presence of \hii\ regions in the RQ sample, which reduces the combined \oiii/\Hb\ ratios with respect to that of pure AGNs. The smaller difference in  $A_{V}^{\rm neb}$ could also be interpreted as due to the removal of objects containing interstellar matter in the mCRQ sample.

We also see that, in the cleaned samples, the difference in the \oiii\ line widths now disappears at large ${\cal R}$. Hence having broader \oiii\ lines only in radio-detected AGNs at $\log {\cal R} < 15.8$ supports the idea that the radio emission in these objects is associated with accretion disk winds (or poorly collimated jets) and that line broadening comes from depositing part of the wind energy into the medium within the narrow line region.

But the most important result here is that the values of CI and $b/a$ are still significantly larger in the CRL than in the mCRQ sample, confirming that there is a real difference in optical morphology between radio-loud and radio-quiet objects matched in AGN parameters. 

\begin{figure*}
\centering
\includegraphics[scale=1]{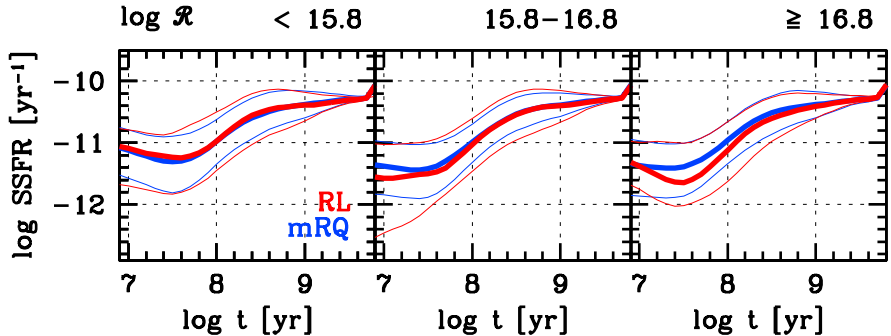}
\caption{Star-formation histories for the RL and the matched RQ
  samples, for the same three bins in ${\cal R}$ as in Fig. \ref{fig:compmatch}. The medians of the specific star-formation rates are the represented by the thick curves and the the 25th and 75th percentiles by the thin ones (red for RL and blue for RQ).
  }
 \label{fig:sfh}
\end{figure*}

We now investigate the differences in star-formation histories of RL and RQ galaxies.  The \starlight\ analysis of SDSS spectra allows us to study the star-formation histories of our galaxies.  

In Fig. 9 we compare the specific star-formation rate (SSFR) of RL and
matched RQ galaxies as a function of lookback time, dividing them as
before into three bins in R of the parent RL objects. The SSFR is
defined as the ratio of the mass converted in stars at a time t to the
total mass ever converted in stars. It is obtained using the \starlight\ 
synthesis modeling which decomposes the stellar populations of each
galaxy into a combination of simple stellar populations of various ages
(\citealp{Asari.etal.2007a}, Sect.\ 4.2 and specifically equation 6). Thus, for
each galaxy, we have a smooth curve of SSFR. What we show in the plot is
the median of all curves for all galaxies in a given $\log {\cal R}$ bin, as well
as the 25th and 75th percentiles.
Both the RL and the mRQ samples have a greater recent SFR in the bin with the smallest ${\cal R}$ of the parent RL galaxy. Since ${\cal R}$ increases with the stellar mass (see Fig. \ref{fig:compmatch}) this may be only an effect of downsizing, where the less massive local galaxies presently form more stars with respect to their mass. 
Significant differences
($p < 0.05$ in the AD and MW tests) occur only
 at $\log t \le
  7.3$ for the $15.8 \le \log {\cal R} < 16.8$ bin, and at $9.7 \le \log t \le
  9.8$ for the $\log {\cal R} \ge 16.8$ bin.
However, the upturn for RL
  at ages $\log \rm t <7.5$ is probably artificial and should be
  disregarded (see the discussion on blue horizontal branches stars in
  stellar population models in \citealp{Ocvirk.2010a} and
  \citealp{Stasinska.etal.2015a}).
 The difference at large ages in the bin $\log {\cal R} \ge 16.8$ is noticable only in the last two age bins.  

  }

\section{Morphological classification}
\label{sec:morph}

Fig. \ref{fig:compmatch} reveals a puzzling picture: RL and RQ galaxies have the same $u-r$ color but differ in concentration index, CI being, on average, larger for RL objects.  This result is a priori surprising, since it is known that galaxy colors and concentration index are correlated. However, there is a certain dispersion in the relation \citep[e.g.][]{Dobrycheva.Melnyk.2012a}.  This is why it is interesting to confirm our conclusion on concentration index by visually inspecting the optical morphologies of the galaxies in our RL and matched RQ samples.

We have selected all of the galaxies from the RL sample that have at least one pair (314 objects) and their closest mRQ match (268 objects, of which 233 are a match to only one RL parent). Six classifiers looked at $102.4\arcsec \times 102.4\arcsec$ color images of a total of 582 galaxies. We show the results for all classifiers.  Our classification scheme labeled galaxies according to (a) their morphology (elliptical, distorted, spiral, lenticular, and ring) and (b) interaction signatures (major or minor merger based on sky projection, tail, suspected interaction, or no sign of interaction).

Mergers are defined as objects having another bright source very close or superimposed on their image; major mergers are those for which the brightness of two interacting objects are comparable by eye. Suspected interactions refer to galaxies with a close---but not too close---companion and a low-surface brightness bridge or some small disturbance in one of the interacting galaxies. Distance issues are minimized by the fact that the RL and mRQ samples are matched in redshift.

\begin{figure}
\centering
\includegraphics[scale=0.5]{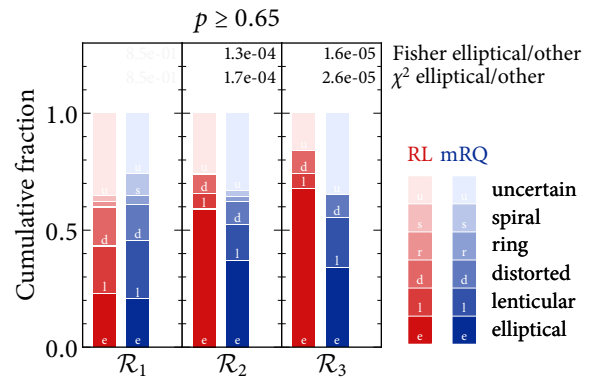}
\caption{Cumulative fraction of RL (shades of red) in three ${\cal R}$
  bins and their matched RQ galaxies (shades of blue), which we 
classified as
  elliptical, lenticular, distorted, ring, or spiral galaxies.  We
    classify galaxies for which $> 65\%$ of the classifiers (four out of
    six) agree; the remaining objects are marked as uncertain. We use the same bins in
  ${\cal R}$ as in Fig. \ref{fig:compmatch}: ${\cal R}_{1}$ corresponds to $\log {\cal R} < 15.8$, ${\cal R}_{2}$ to $15.8 \le \log {\cal R} < 16.8$, and ${\cal R}_3$ to $\log {\cal R} \ge 16.8$; there are 84, 150, and
    80 RL galaxies in ${\cal R}_{1}$, ${\cal R}_{2}$ and ${\cal R}_3$
  respectively, and there is the same number of paired RQ galaxies. The
  values on the top of each panel show the results of Fisher and
  $\chi^2$ two-sample tests for elliptical and other galaxies.  As
  in Fig. \ref{fig:compmatch}, results more significant than $p <
    0.003$ are marked in black, more significant than $p <
    0.05$ in dark gray.
  For clarity, fractions that contribute very little to the
    total have been adequately colored but not labeled with a letter.
}
 \label{fig:morph}
\end{figure}

\begin{figure}
\centering
\includegraphics[scale=0.5]{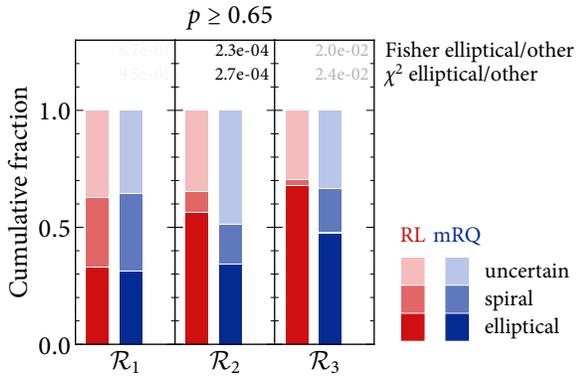}
\caption{Results on morphology from Galaxy Zoo 1: cumulative fraction of RL (red) in three ${\cal R}$ bins and their matched RQ galaxies (blue) classified in GZ1 as elliptical, spiral, or uncertain. The layout is the same as in Fig. \ref{fig:morph}.}
 \label{fig:morphGZ}
\end{figure}

The panels of Fig. \ref{fig:morph} show in different shades of red the fraction of RL objects classified as elliptical, lenticular, distorted, ring, and spiral galaxies for the same radio loudness bins as Fig.  \ref{fig:compmatch} for all classifiers. Galaxies are classified as belonging to a morphological class if  the agreement between classifiers is greater than 65\%
 (i.e.\ at least four out of six classifiers agree). Otherwise, galaxies are marked as uncertain. RQ objects matched to the RL in each ${\cal R}$ bin are shown in shades of blue. The behavior of the elliptical fraction closely mimics that of CI: the fraction of ellipticals among RL objects is larger than among the matched RQ galaxies ($p < 0.003$) for ${\cal R} > 15.8$. We note that the low-CI galaxies are not spirals: in our classification scheme, they are either lenticulars/S0 systems or distorted galaxies. The fraction of RL ellipticals is larger in the bins with higher ${\cal R}$, while, as expected, there is no significant trend in the fraction of ellipticals in the matched RQ sample.

As a comparison, we also show results for Galaxy Zoo 1 (GZ1, \citealp{Lintott.etal.2008a, Lintott.etal.2011a}). In the Galaxy Zoo project, volunteers morphologically classified almost 900,000 of SDSS galaxies. In the GZ1, the classifiers could choose between six categories: elliptical galaxy, clockwise, or anticlockwise spiral galaxy, edge-on galaxy, star/artifact or merger. About 93\% of our
 RL and first matches from RQ galaxies were classified in this project. Each galaxy was classified repeatedly by different classifiers and for each of them the GZ1 gives the probability of the assigned morphological type. Fig.  \ref{fig:morphGZ} shows the fractions of elliptical, spiral, and uncertain galaxies, including only objects for which the probabilities of a galaxy being spiral or elliptical are at least 65\% (which should result in $< 10\%$ of misclassifications, according to \citealp{Lintott.etal.2008a}).  Although our and GZ1 classification schemes are different, the results on the fractions of ellipticals in the RL and matched RQ samples are consistent.
 
\begin{figure}
\includegraphics[scale=0.5]{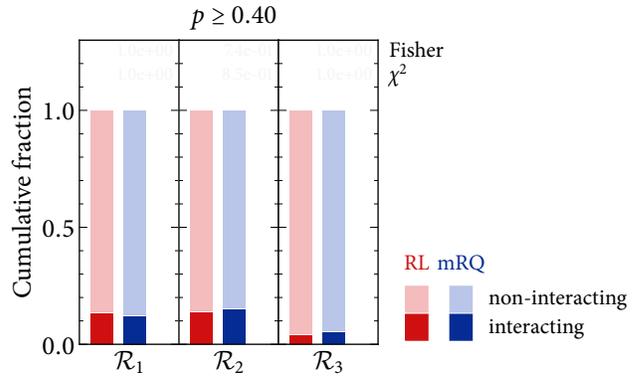}
\caption{Cumulative fraction of RL galaxies in three ${\cal R}$ bins and their matched RQ galaxies showing signs of interaction. We classify galaxies as interacting if $> 40\%$ of the classifiers ($>2$ out of 6) have seen clear signs of interactions. The layout and the number of galaxies in each bin is the same as for Fig.\ \ref{fig:morph}. }
 \label{fig:inter}
\end{figure}

\begin{figure}
\includegraphics[scale=0.5]{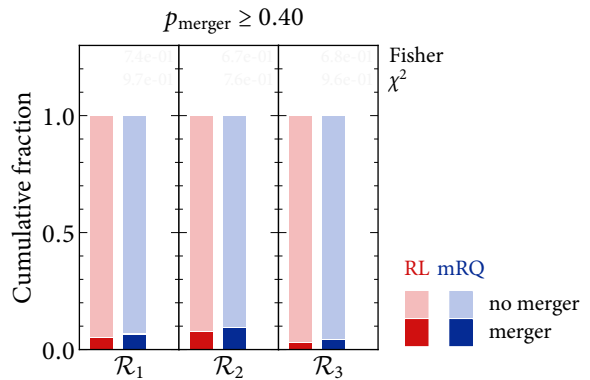}
\caption{Cumulative fraction of RL galaxies in three ${\cal R}$ bins and their matched RQ galaxies classified as mergers in GZ1. The panels show the fraction of galaxies with probability of being mergers larger than 40\%. The layout is the same as for Fig.\ \ref{fig:morphGZ}. }
 \label{fig:interGZ}
\end{figure}

Fig.\ \ref{fig:inter} and \ref{fig:interGZ} show the fraction of signatures of interaction for the RL and mRQ samples in GZ1 and in our classification schemes.  In the case of our results, we use a conservative definition of interactions and consider as interacting only galaxies with a tail, or major and minor mergers. After \cite{Darg.etal.2010a}, we qualify as interacting, in both schemes, all the galaxies for which the probability of being merger (i.e. for which the fraction of votes for being a merger) is larger or equal to 40\%.

There is no
systematic difference between RL and RQ galaxies in either panel.  This result is in apparent contradiction with \citet{Chiaberge.etal.2015a}, who find that radio-loud galaxies are mergers. However, their sample is quite different from ours: they consider FR II radio galaxies at redshifts $z> 1$, while our sample contains mostly compact radio sources at $z \sim 0.1$. In addition, they did not compare their radio-loud and radio-quiet samples in mass bins.  Since on average radio-loud galaxies have higher masses than radio-quiet ones, and since more massive galaxies tend to be found in the densest environments \citep{Goto.etal.2003a}, their result might well be related to galaxy masses rather than to radio loudness.  The larger fraction of mergers in the radio-loud sample of \citet{Chiaberge.etal.2015a} can also be related to higher BH masses. \citet{Chiaberge.etal.2015a} do not provide information on BH masses in their samples, but assuming that their radio galaxies act at similar Eddington ratios as ours, larger masses of their BHs ($\sim 10^9$\msun) might be deduced from the larger luminosities of their hosts or their larger radio luminosities.

We must also note that one cannot exclude the possibility that many (if not most) very radio-loud AGNs are triggered by mergers. However, they do not have to be major mergers.  It may
 be that the activation of the quasar phase took place more than $10^8$ years after the merger event \citep[see, e.g.][and references therein]{Blank.Duschl.2016a}.  Then, the signs of a galaxy merger at the epoch of the AGN activation will be visible only in the very central region, particularly if it is a minor merger.  For example, the fact that Cyg A, a nearby ($z=0.056$) powerful FR II radio galaxy, was most likely activated by a merger, was deduced only after using adaptive optics imaging \citep{Canalizo.etal.2003a,Privon.etal.2012a}.

A few notes on the statistical tests used in this section. We have
  considered the morphological classifications as categorical data tables. Thus we confront the null hypothesis that the RL and mRQ have the same morphological classifications by applying the Fisher exact test and the chi-square two-sample test. The Fisher exact test can be applied only for binary data, so the null hypothesis is that RL and mRQ have the same proportion of ellipticals {and other classes)}, or the same proportion of interacting and non-interacting galaxies. The chi-square two-sample test is applicable to categorical data in contingency tables, and it is reliable when most ($\sim 80\%$) data bins have $> 5$ counts. To meet this criterion, we define our null hypothesis as RL and mRQ having the same distribution of ellipticals and others. Because we want to avoid choosing a model for our data and for data outliers, we apply non-parametric tests. Since non-parametric tests are unknown to exist in the Bayesian framework (see, e.g. \citealp{Wall.Jenkins.2003a}, chapter 5), we are obliged to rely on classical rather than Bayesian statistical methods.

\section{Conclusions}

\label{sec:conclus}

In order to find some clues regarding the conditions leading to the presence of radio jets in AGNs,  we have compared samples of RL and RQ AGNs. The AGNs were extracted from the SDSS DR7 database, excluding those objects for which the \Ha\ luminosity could be 
dominated by the ionization  by old stellar populations \citep{CidFernandes.etal.2011a}. The RL sample was extracted from the catalog of \cite{Best.Heckman.2012a} removing those galaxies where according to those authors the radio emission comes from star formation, and removing a few misidentifications.

In order to avoid effects of having different average BH masses and accretion rates in radio-loud and radio-quiet samples of AGNs, we compared properties of their hosts by matching them in BH mass and Eddington ratio. Matching in both these parameters allows us to study the differences between galaxies hosting RL and RQ AGNs that have the same basic accretion parameters.

We have limited our AGN samples to Type 2, i.e. obscured nuclei, so as to avoid pollution of the spectra by the AGN continuum and by features from the broad line region. We have also limited them to $\lambda > 0.003$, to focus mainly on radiatively efficient accretion flows (\citealp[e.g.][]{Best.Heckman.2012a, Stern.Laor.2013a}).
  
We have noted that delimiting the AGN sample in the BPT diagram by the commonly used K01 line, which is generally considered to delimit pure AGNs from   composite ones, the presence of a certain amount of star formation affects some of our results. We have therefore also considered `cleaned' samples, by removing objects that lie not far
above the K01 line. This results in a sample that is much less affected by star formation, but unfortunately reduced. However, we find that the results from the cleaned samples corroborate the ones from the uncleaned ones.
      
Our main results can be summarized as follows.
\begin{enumerate}
\item The colors $u-r$ of the RL hosts and of their paired RQ hosts are similar and do not show any dependence on ${\cal R}$.
\item We do not find any significant difference between the  RL and RQ samples in star-formation histories  deduced from the stellar population analysis.
\item The RL AGNs at $\log {\cal R} < 15.8$ have broader \oiii\ lines, which may be connected to the fact that their radio emission comes from accretion disk winds.
\item The concentration index CI of the RL hosts is larger than that of the paired RQ hosts
for a radio loudness log ${\cal R} > 15.8$, and so is the geometrical parameter $b/a$.
\item The RL galaxies are of earlier morphological types (mostly elliptical)  than the RQ galaxies (mostly lenticular).
\item The fractions of interaction signatures in RL and RQ host galaxies are similar.
\end{enumerate}
 It must be noted that even as regards parameters for which the RL and RQ samples show differences, there is 
a large overlap between the two samples. 
However, if the jet production is associated with one of the two accretion modes suggested to interchangeably operate at similar rates (\citealp[][]{Kording.Jester.Fender.2006a}; see also \citealp{Livio.Pringle.King.2003a, Nipoti.Blundell.Binney.2005a}), the overlap could be caused by the fact that some galaxies in RQ sample can be in fact RL but in a radio quiescence state at the moment. If present, this mixing cannot be removed from our sample, and may cause the results to be blurry.

Nevertheless, our results demonstrate that the efficiency of the jet production is not fully determined by just the Eddington ratio and BH mass. From a theoretical point of view, further differentiation  of the jet production efficiency is likely to be provided by the magnetic flux accumulated in  AGN centers  and, in the case of launching a jet via the Blandford-Znajek mechanism, also by the BH spin.  

The relation between radio loudness and the host-galaxy morphological type that we have found (the larger the radio loudness, the earlier the type of the host galaxy) is actually consistent with these theoretical predictions. A recent study by \cite{Ruiz.etal.2015a} suggests that heavier dark matter halos are expected to have a larger amount of gas, the inflow of this gas to galactic centers may provide the right conditions to advect and accumulate the magnetic flux there \citep{Cao.2011a}. This may result in the RL AGN pre-phase suggested by \cite{Sikora.etal.2013a} and explored by \cite{Sikora.Begelman.2013a}. In addition, as modeling of cosmological evolution of supermassive BHs indicates, the BHs in gas-poor galaxies tend to have larger spins than the BHs hosted by later-type galaxies \citep{Volonteri.etal.2013a}.

\section*{ACKNOWLEDGMENTS}

The authors thank the anonymous referee and the AAS Journal statistician for the useful comments, \L ukasz Stawarz and Fabio R. Herpich for fruitful discussions, and Rafael Pacheco Cardoso and Graciana Brum for their help in the morphological classification. 
This work was carried out within the framework of the Polish National Science Centre grant UMO-2013/09/B/ST9/00026.  Gra\.zyna Stasi\'nska was partially supported by the National Research Centre, Poland, DEC-2013/08/M/ST9/00664, within the framework of the HECOLS International Associated Laboratory.  D.K.W was partly supported by the Polish National Science Centre grant UMO-2013/09/B/ST9/00599.  G.S. and N.V.A. acknowledge the support from the CAPES CSF--PVE project 88881.068116/2014-01. The Sloan Digital Sky Survey is a joint project of The University of Chicago, Fermilab, the Institute for Advanced Study, the Japan Participation Group, the Johns Hopkins University, the Los Alamos National Laboratory, the Max-Planck-Institute for Astronomy, the Max-Planck-Institute for Astrophysics, New Mexico State University, Princeton University, the United States Naval Observatory, and the University of Washington.  Funding for the project has been provided by the Alfred P. Sloan Foundation, the Participating Institutions, the National Aeronautics and Space Administration, the National Science Foundation, the U.S. Department of Energy, the Japanese Monbukagakusho, and the Max Planck Society.

\bibliography{references}

\begin{thebibliography}{}
\expandafter\ifx\csname natexlab\endcsname\relax\def\natexlab#1{#1}\fi

\bibitem[{{Abazajian} {et~al.}(2009){Abazajian}, {Adelman-McCarthy},
  {Ag{\"u}eros}, {Allam}, {Allende Prieto}, {An}, {Anderson}, {Anderson},
  {Annis}, {Bahcall}, \& et~al.}]{Abazajian.etal.2009a}
{Abazajian}, K.~N., {Adelman-McCarthy}, J.~K., {Ag{\"u}eros}, M.~A., {et~al.}
  2009, \apjs, 182, 543

\bibitem[{{Asari} {et~al.}(2007){Asari}, {Cid Fernandes}, {Stasi{\'n}ska},
  {Torres-Papaqui}, {Mateus}, {Sodr{\'e}}, {Schoenell}, \&
  {Gomes}}]{Asari.etal.2007a}
{Asari}, N.~V., {Cid Fernandes}, R., {Stasi{\'n}ska}, G., {et~al.} 2007,
  \mnras, 381, 263

\bibitem[{{Baldwin} {et~al.}(1981){Baldwin}, {Phillips}, \&
  {Terlevich}}]{Baldwin.Phillips.Terlevich.1981a}
{Baldwin}, J.~A., {Phillips}, M.~M., \& {Terlevich}, R. 1981, \pasp, 93, 5

\bibitem[{{Becker} {et~al.}(1994){Becker}, {White}, {Helfand}, {Greeg}, \&
  {Perley}}]{Becker.etal.1994a}
{Becker}, R.~H., {White}, R.~L., {Helfand}, D.~J., {Greeg}, M.~D., \& {Perley},
  R.~A. 1994, in Bulletin of the American Astronomical Society, Vol.~26,
  American Astronomical Society Meeting Abstracts, 1317

\bibitem[{{Bertelli} {et~al.}(1994){Bertelli}, {Bressan}, {Chiosi}, {Fagotto},
  \& {Nasi}}]{Bertelli.etal.1994a}
{Bertelli}, G., {Bressan}, A., {Chiosi}, C., {Fagotto}, F., \& {Nasi}, E. 1994,
  \aaps, 106, 275

\bibitem[{{Bessiere} {et~al.}(2012){Bessiere}, {Tadhunter}, {Ramos Almeida}, \&
  {Villar Mart{\'{\i}}n}}]{Bessiere.etal.2012a}
{Bessiere}, P.~S., {Tadhunter}, C.~N., {Ramos Almeida}, C., \& {Villar
  Mart{\'{\i}}n}, M. 2012, \mnras, 426, 276

\bibitem[{{Best} \& {Heckman}(2012)}]{Best.Heckman.2012a}
{Best}, P.~N., \& {Heckman}, T.~M. 2012, \mnras, 421, 1569

\bibitem[{{Best} {et~al.}(2005){Best}, {Kauffmann}, {Heckman}, {Brinchmann},
  {Charlot}, {Ivezi{\'c}}, \& {White}}]{Best.etal.2005a}
{Best}, P.~N., {Kauffmann}, G., {Heckman}, T.~M., {et~al.} 2005, \mnras, 362,
  25

\bibitem[{{Blandford} \& {Znajek}(1977)}]{1977MNRAS.179..433B}
{Blandford}, R.~D., \& {Znajek}, R.~L. 1977, \mnras, 179, 433

\bibitem[{{Blank} \& {Duschl}(2016)}]{Blank.Duschl.2016a}
{Blank}, M., \& {Duschl}, W.~J. 2016, \mnras, 462, 2246

\bibitem[{{Blundell} \& {Kuncic}(2007)}]{Blundell.Kuncic.2007a}
{Blundell}, K.~M., \& {Kuncic}, Z. 2007, \apjl, 668, L103

\bibitem[{{Bruzual} \& {Charlot}(2003)}]{Bruzual.Charlot.2003a}
{Bruzual}, G., \& {Charlot}, S. 2003, \mnras, 344, 1000

\bibitem[{{Canalizo} {et~al.}(2003){Canalizo}, {Max}, {Whysong}, {Antonucci},
  \& {Dahm}}]{Canalizo.etal.2003a}
{Canalizo}, G., {Max}, C., {Whysong}, D., {Antonucci}, R., \& {Dahm}, S.~E.
  2003, \apj, 597, 823

\bibitem[{{Cao}(2011)}]{Cao.2011a}
{Cao}, X. 2011, \apj, 737, 94

\bibitem[{{Cardelli} {et~al.}(1989){Cardelli}, {Clayton}, \&
  {Mathis}}]{Cardelli.Clayton.Mathis.1989a}
{Cardelli}, J.~A., {Clayton}, G.~C., \& {Mathis}, J.~S. 1989, \apj, 345, 245

\bibitem[{{Chabrier}(2003)}]{Chabrier.2003a}
{Chabrier}, G. 2003, \pasp, 115, 763

\bibitem[{{Chiaberge} {et~al.}(2015){Chiaberge}, {Gilli}, {Lotz}, \&
  {Norman}}]{Chiaberge.etal.2015a}
{Chiaberge}, M., {Gilli}, R., {Lotz}, J.~M., \& {Norman}, C. 2015, \apj, 806,
  147

\bibitem[{{Cid Fernandes} {et~al.}(2005){Cid Fernandes}, {Mateus}, {Sodr{\'e}},
  {Stasi{\'n}ska}, \& {Gomes}}]{CidFernandes.etal.2005a}
{Cid Fernandes}, R., {Mateus}, A., {Sodr{\'e}}, L., {Stasi{\'n}ska}, G., \&
  {Gomes}, J.~M. 2005, \mnras, 358, 363

\bibitem[{{Cid Fernandes} {et~al.}(2011){Cid Fernandes}, {Stasi{\'n}ska},
  {Mateus}, \& {Vale Asari}}]{CidFernandes.etal.2011a}
{Cid Fernandes}, R., {Stasi{\'n}ska}, G., {Mateus}, A., \& {Vale Asari}, N.
  2011, \mnras, 413, 1687

\bibitem[{{Cid Fernandes} {et~al.}(2010){Cid Fernandes}, {Stasi{\'n}ska},
  {Schlickmann}, {Mateus}, {Vale Asari}, {Schoenell}, \&
  {Sodr{\'e}}}]{CidFernandes.etal.2010a}
{Cid Fernandes}, R., {Stasi{\'n}ska}, G., {Schlickmann}, M.~S., {et~al.} 2010,
  \mnras, 403, 1036

\bibitem[{{Cid Fernandes} {et~al.}(2009){Cid Fernandes}, {Schoenell}, {Gomes},
  {Asari}, {Schlickmann}, {Mateus}, {Stasinska}, {Sodr{\'e}}, {Torres-Papaqui},
  \& {Seagal Collaboration}}]{CidFernandes.etal.2009a}
{Cid Fernandes}, R., {Schoenell}, W., {Gomes}, J.~M., {et~al.} 2009, in Revista
  Mexicana de Astronomia y Astrofisica Conference Series, Vol.~35, Revista
  Mexicana de Astronomia y Astrofisica Conference Series, 127--132

\bibitem[{{Condon} {et~al.}(1998){Condon}, {Cotton}, {Greisen}, {Yin},
  {Perley}, {Taylor}, \& {Broderick}}]{Condon.etal.1998}
{Condon}, J.~J., {Cotton}, W.~D., {Greisen}, E.~W., {et~al.} 1998, \aj, 115,
  1693

\bibitem[{{Darg} {et~al.}(2010){Darg}, {Kaviraj}, {Lintott}, {Schawinski},
  {Sarzi}, {Bamford}, {Silk}, {Andreescu}, {Murray}, {Nichol}, {Raddick},
  {Slosar}, {Szalay}, {Thomas}, \& {Vandenberg}}]{Darg.etal.2010a}
{Darg}, D.~W., {Kaviraj}, S., {Lintott}, C.~J., {et~al.} 2010, \mnras, 401,
  1552

\bibitem[{{Dicken} {et~al.}(2012){Dicken}, {Tadhunter}, {Axon}, {Morganti},
  {Robinson}, {Kouwenhoven}, {Spoon}, {Kharb}, {Inskip}, {Holt}, {Ramos
  Almeida}, \& {Nesvadba}}]{Dicken.etal.2012a}
{Dicken}, D., {Tadhunter}, C., {Axon}, D., {et~al.} 2012, \apj, 745, 172

\bibitem[{{Dobrycheva} \& {Melnyk}(2012)}]{Dobrycheva.Melnyk.2012a}
{Dobrycheva}, D., \& {Melnyk}, O. 2012, Advances in Astronomy and Space
  Physics, 2, 42

\bibitem[{{Donoso} {et~al.}(2010){Donoso}, {Li}, {Kauffmann}, {Best}, \&
  {Heckman}}]{Donoso.etal.2010a}
{Donoso}, E., {Li}, C., {Kauffmann}, G., {Best}, P.~N., \& {Heckman}, T.~M.
  2010, \mnras, 407, 1078

\bibitem[{{Dunlop} {et~al.}(2003){Dunlop}, {McLure}, {Kukula}, {Baum}, {O'Dea},
  \& {Hughes}}]{Dunlop.etal.2003a}
{Dunlop}, J.~S., {McLure}, R.~J., {Kukula}, M.~J., {et~al.} 2003, \mnras, 340,
  1095

\bibitem[{{Eisenstein} {et~al.}(2001){Eisenstein}, {Annis}, {Gunn}, {Szalay},
  {Connolly}, {Nichol}, {Bahcall}, {Bernardi}, {Burles}, {Castander},
  {Fukugita}, {Hogg}, {Ivezi{\'c}}, {Knapp}, {Lupton}, {Narayanan}, {Postman},
  {Reichart}, {Richmond}, {Schneider}, {Schlegel}, {Strauss}, {SubbaRao},
  {Tucker}, {Vanden Berk}, {Vogeley}, {Weinberg}, \&
  {Yanny}}]{Eisenstein.etal.2001a}
{Eisenstein}, D.~J., {Annis}, J., {Gunn}, J.~E., {et~al.} 2001, \aj, 122, 2267

\bibitem[{{Ekholm} {et~al.}(2001){Ekholm}, {Baryshev}, {Teerikorpi}, {Hanski},
  \& {Paturel}}]{Ekholm.etal.2001a}
{Ekholm}, T., {Baryshev}, Y., {Teerikorpi}, P., {Hanski}, M.~O., \& {Paturel},
  G. 2001, \aap, 368, L17

\bibitem[{{Falder} {et~al.}(2010){Falder}, {Stevens}, {Jarvis}, {Hardcastle},
  {Lacy}, {McLure}, {Hatziminaoglou}, {Page}, \&
  {Richards}}]{Falder.etal.2010a}
{Falder}, J.~T., {Stevens}, J.~A., {Jarvis}, M.~J., {et~al.} 2010, \mnras, 405,
  347

\bibitem[{{Fanaroff} \& {Riley}(1974)}]{Fanaroff.Riley.1974a}
{Fanaroff}, B.~L., \& {Riley}, J.~M. 1974, \mnras, 167, 31P

\bibitem[{{Ferrarese} \& {Merritt}(2000)}]{Ferrarese.Merritt.2000a}
{Ferrarese}, L., \& {Merritt}, D. 2000, \apjl, 539, L9

\bibitem[{{Floyd} {et~al.}(2013){Floyd}, {Dunlop}, {Kukula}, {Brown}, {McLure},
  {Baum}, \& {O'Dea}}]{Floyd.etal.2013a}
{Floyd}, D.~J.~E., {Dunlop}, J.~S., {Kukula}, M.~J., {et~al.} 2013, \mnras,
  429, 2

\bibitem[{{Floyd} {et~al.}(2004){Floyd}, {Kukula}, {Dunlop}, {McLure},
  {Miller}, {Percival}, {Baum}, \& {O'Dea}}]{Floyd.etal.2004a}
{Floyd}, D.~J.~E., {Kukula}, M.~J., {Dunlop}, J.~S., {et~al.} 2004, \mnras,
  355, 196

\bibitem[{{Gadotti} \& {Kauffmann}(2009)}]{Gadotti.Kauffmann.2009a}
{Gadotti}, D.~A., \& {Kauffmann}, G. 2009, \mnras, 399, 621

\bibitem[{{Goto} {et~al.}(2003){Goto}, {Yamauchi}, {Fujita}, {Okamura},
  {Sekiguchi}, {Smail}, {Bernardi}, \& {Gomez}}]{Goto.etal.2003a}
{Goto}, T., {Yamauchi}, C., {Fujita}, Y., {et~al.} 2003, \mnras, 346, 601

\bibitem[{{G{\"u}rkan} {et~al.}(2015){G{\"u}rkan}, {Hardcastle}, {Jarvis},
  {Smith}, {Bourne}, {Dunne}, {Maddox}, {Ivison}, \&
  {Fritz}}]{Gurkan.etal.2015a}
{G{\"u}rkan}, G., {Hardcastle}, M.~J., {Jarvis}, M.~J., {et~al.} 2015, \mnras,
  452, 3776

\bibitem[{{Heckman} \& {Best}(2014)}]{Heckman.Best.2014a}
{Heckman}, T.~M., \& {Best}, P.~N. 2014, \araa, 52, 589

\bibitem[{{Kauffmann} {et~al.}(2008){Kauffmann}, {Heckman}, \&
  {Best}}]{Kauffmann.Heckman.Best.2008a}
{Kauffmann}, G., {Heckman}, T.~M., \& {Best}, P.~N. 2008, \mnras, 384, 953

\bibitem[{{Kauffmann} {et~al.}(2003){Kauffmann}, {Heckman}, {Tremonti},
  {Brinchmann}, {Charlot}, {White}, {Ridgway}, {Brinkmann}, {Fukugita}, {Hall},
  {Ivezi{\'c}}, {Richards}, \& {Schneider}}]{Kauffmann.etal.2003c}
{Kauffmann}, G., {Heckman}, T.~M., {Tremonti}, C., {et~al.} 2003, \mnras, 346,
  1055

\bibitem[{{Kellermann} {et~al.}(2016){Kellermann}, {Condon}, {Kimball},
  {Perley}, \& {Ivezic}}]{Kellermann.etal.2016a}
{Kellermann}, K.~I., {Condon}, J.~J., {Kimball}, A.~E., {Perley}, R.~A., \&
  {Ivezic}, Z. 2016, ArXiv e-prints, arXiv:1608.04586

\bibitem[{{Kellermann} {et~al.}(1989){Kellermann}, {Sramek}, {Schmidt},
  {Shaffer}, \& {Green}}]{Kellermann.etal.1989a}
{Kellermann}, K.~I., {Sramek}, R., {Schmidt}, M., {Shaffer}, D.~B., \& {Green},
  R. 1989, \aj, 98, 1195

\bibitem[{{Kewley} {et~al.}(2001){Kewley}, {Dopita}, {Sutherland}, {Heisler},
  \& {Trevena}}]{Kewley.etal.2001a}
{Kewley}, L.~J., {Dopita}, M.~A., {Sutherland}, R.~S., {Heisler}, C.~A., \&
  {Trevena}, J. 2001, \apj, 556, 121

\bibitem[{{Kimball} {et~al.}(2011){Kimball}, {Kellermann}, {Condon},
  {Ivezi{\'c}}, \& {Perley}}]{Kimball.etal.2011a}
{Kimball}, A.~E., {Kellermann}, K.~I., {Condon}, J.~J., {Ivezi{\'c}}, {\v Z}.,
  \& {Perley}, R.~A. 2011, \apjl, 739, L29

\bibitem[{{K{\"o}rding} {et~al.}(2006){K{\"o}rding}, {Jester}, \&
  {Fender}}]{Kording.Jester.Fender.2006a}
{K{\"o}rding}, E.~G., {Jester}, S., \& {Fender}, R. 2006, \mnras, 372, 1366

\bibitem[{{Kormendy} \& {Ho}(2013)}]{Kormendy.Ho.2013a}
{Kormendy}, J., \& {Ho}, L.~C. 2013, \araa, 51, 511

\bibitem[{{Lal} \& {Ho}(2010)}]{Lal.Ho.2010a}
{Lal}, D.~V., \& {Ho}, L.~C. 2010, \aj, 139, 1089

\bibitem[{{Laor}(2000)}]{Laor2000a}
{Laor}, A. 2000, \apjl, 543, L111

\bibitem[{{Le Borgne} {et~al.}(2003){Le Borgne}, {Bruzual}, {Pell{\'o}},
  {Lan{\c c}on}, {Rocca-Volmerange}, {Sanahuja}, {Schaerer}, {Soubiran}, \&
  {V{\'{\i}}lchez-G{\'o}mez}}]{LeBorgne.etal.2003a}
{Le Borgne}, J.-F., {Bruzual}, G., {Pell{\'o}}, R., {et~al.} 2003, \aap, 402,
  433

\bibitem[{{Lintott} {et~al.}(2011){Lintott}, {Schawinski}, {Bamford}, {Slosar},
  {Land}, {Thomas}, {Edmondson}, {Masters}, {Nichol}, {Raddick}, {Szalay},
  {Andreescu}, {Murray}, \& {Vandenberg}}]{Lintott.etal.2011a}
{Lintott}, C., {Schawinski}, K., {Bamford}, S., {et~al.} 2011, \mnras, 410, 166

\bibitem[{{Lintott} {et~al.}(2008){Lintott}, {Schawinski}, {Slosar}, {Land},
  {Bamford}, {Thomas}, {Raddick}, {Nichol}, {Szalay}, {Andreescu}, {Murray}, \&
  {Vandenberg}}]{Lintott.etal.2008a}
{Lintott}, C.~J., {Schawinski}, K., {Slosar}, A., {et~al.} 2008, \mnras, 389,
  1179

\bibitem[{{Livio} {et~al.}(2003){Livio}, {Pringle}, \&
  {King}}]{Livio.Pringle.King.2003a}
{Livio}, M., {Pringle}, J.~E., \& {King}, A.~R. 2003, \apj, 593, 184

\bibitem[{{Lu} {et~al.}(2007){Lu}, {Wang}, {Zhou}, \& {Wu}}]{Lu.etal.2007a}
{Lu}, Y., {Wang}, T., {Zhou}, H., \& {Wu}, J. 2007, \aj, 133, 1615

\bibitem[{{Mandelbaum} {et~al.}(2009){Mandelbaum}, {Li}, {Kauffmann}, \&
  {White}}]{Mandelbaum.etal.2009a}
{Mandelbaum}, R., {Li}, C., {Kauffmann}, G., \& {White}, S.~D.~M. 2009, \mnras,
  393, 377

\bibitem[{{Masters} {et~al.}(2010){Masters}, {Mosleh}, {Romer}, {Nichol},
  {Bamford}, {Schawinski}, {Lintott}, {Andreescu}, {Campbell}, {Crowcroft},
  {Doyle}, {Edmondson}, {Murray}, {Raddick}, {Slosar}, {Szalay}, \&
  {Vandenberg}}]{Masters.etal.2010a}
{Masters}, K.~L., {Mosleh}, M., {Romer}, A.~K., {et~al.} 2010, \mnras, 405, 783

\bibitem[{{Mateus} {et~al.}(2006){Mateus}, {Sodr{\'e}}, {Cid Fernandes},
  {Stasi{\'n}ska}, {Schoenell}, \& {Gomes}}]{Mateus.etal.2006a}
{Mateus}, A., {Sodr{\'e}}, L., {Cid Fernandes}, R., {et~al.} 2006, \mnras, 370,
  721

\bibitem[{{McLure} \& {Jarvis}(2004)}]{McLure.Jarvis.2004a}
{McLure}, R.~J., \& {Jarvis}, M.~J. 2004, \mnras, 353, L45

\bibitem[{{Netzer}(2009)}]{Netzer.2009a}
{Netzer}, H. 2009, \mnras, 399, 1907

\bibitem[{{Nipoti} {et~al.}(2005){Nipoti}, {Blundell}, \&
  {Binney}}]{Nipoti.Blundell.Binney.2005a}
{Nipoti}, C., {Blundell}, K.~M., \& {Binney}, J. 2005, \mnras, 361, 633

\bibitem[{{Ocvirk}(2010)}]{Ocvirk.2010a}
{Ocvirk}, P. 2010, \apj, 709, 88

\bibitem[{{Privon} {et~al.}(2012){Privon}, {Baum}, {O'Dea}, {Gallimore},
  {Noel-Storr}, {Axon}, \& {Robinson}}]{Privon.etal.2012a}
{Privon}, G.~C., {Baum}, S.~A., {O'Dea}, C.~P., {et~al.} 2012, \apj, 747, 46

\bibitem[{{Ramos Almeida} {et~al.}(2013){Ramos Almeida}, {Bessiere},
  {Tadhunter}, {Inskip}, {Morganti}, {Dicken}, {Gonz{\'a}lez-Serrano}, \&
  {Holt}}]{RamosAlmeida.etal.2013a}
{Ramos Almeida}, C., {Bessiere}, P.~S., {Tadhunter}, C.~N., {et~al.} 2013,
  \mnras, 436, 997

\bibitem[{{Ruiz} {et~al.}(2015){Ruiz}, {Trujillo}, \&
  {M{\'a}rmol-Queralt{\'o}}}]{Ruiz.etal.2015a}
{Ruiz}, P., {Trujillo}, I., \& {M{\'a}rmol-Queralt{\'o}}, E. 2015, \mnras, 454,
  1605

\bibitem[{{Runnoe} {et~al.}(2012){Runnoe}, {Brotherton}, \&
  {Shang}}]{2012MNRAS.422..478R}
{Runnoe}, J.~C., {Brotherton}, M.~S., \& {Shang}, Z. 2012, \mnras, 422, 478

\bibitem[{{Shen} {et~al.}(2009){Shen}, {Strauss}, {Ross}, {Hall}, {Lin},
  {Richards}, {Schneider}, {Weinberg}, {Connolly}, {Fan}, {Hennawi}, {Shankar},
  {Vanden Berk}, {Bahcall}, \& {Brunner}}]{Shen.etal.2009a}
{Shen}, Y., {Strauss}, M.~A., {Ross}, N.~P., {et~al.} 2009, \apj, 697, 1656

\bibitem[{{Sikora} \& {Begelman}(2013)}]{Sikora.Begelman.2013a}
{Sikora}, M., \& {Begelman}, M.~C. 2013, \apjl, 764, L24

\bibitem[{{Sikora} {et~al.}(2013){Sikora}, {Stasi{\'n}ska},
  {Kozie{\l}-Wierzbowska}, {Madejski}, \& {Asari}}]{Sikora.etal.2013a}
{Sikora}, M., {Stasi{\'n}ska}, G., {Kozie{\l}-Wierzbowska}, D., {Madejski},
  G.~M., \& {Asari}, N.~V. 2013, \apj, 765, 62

\bibitem[{{Sikora} {et~al.}(2007){Sikora}, {Stawarz}, \&
  {Lasota}}]{Sikora.Stawarz.Lasota.2007a}
{Sikora}, M., {Stawarz}, {\L}., \& {Lasota}, J.-P. 2007, \apj, 658, 815

\bibitem[{{Stasi{\'n}ska} {et~al.}(2006){Stasi{\'n}ska}, {Cid Fernandes},
  {Mateus}, {Sodr{\'e}}, \& {Asari}}]{Stasinska.etal.2006a}
{Stasi{\'n}ska}, G., {Cid Fernandes}, R., {Mateus}, A., {Sodr{\'e}}, L., \&
  {Asari}, N.~V. 2006, \mnras, 371, 972

\bibitem[{{Stasi{\'n}ska} {et~al.}(2015){Stasi{\'n}ska}, {Costa-Duarte}, {Vale
  Asari}, {Cid Fernandes}, \& {Sodr{\'e}}}]{Stasinska.etal.2015a}
{Stasi{\'n}ska}, G., {Costa-Duarte}, M.~V., {Vale Asari}, N., {Cid Fernandes},
  R., \& {Sodr{\'e}}, L. 2015, \mnras, 449, 559

\bibitem[{{Stern} \& {Laor}(2013)}]{Stern.Laor.2013a}
{Stern}, J., \& {Laor}, A. 2013, \mnras, 431, 836

\bibitem[{{Strauss} {et~al.}(2002){Strauss}, {Weinberg}, {Lupton}, {Narayanan},
  {Annis}, {Bernardi}, {Blanton}, {Burles}, {Connolly}, {Dalcanton}, {Doi},
  {Eisenstein}, {Frieman}, {Fukugita}, {Gunn}, {Ivezi{\'c}}, {Kent}, {Kim},
  {Knapp}, {Kron}, {Munn}, {Newberg}, {Nichol}, {Okamura}, {Quinn}, {Richmond},
  {Schlegel}, {Shimasaku}, {SubbaRao}, {Szalay}, {Vanden Berk}, {Vogeley},
  {Yanny}, {Yasuda}, {York}, \& {Zehavi}}]{Strauss.etal.2002a}
{Strauss}, M.~A., {Weinberg}, D.~H., {Lupton}, R.~H., {et~al.} 2002, \aj, 124,
  1810

\bibitem[{{Thomas} {et~al.}(2013){Thomas}, {Steele}, {Maraston}, {Johansson},
  {Beifiori}, {Pforr}, {Str{\"o}mb{\"a}ck}, {Tremonti}, {Wake}, {Bizyaev},
  {Bolton}, {Brewington}, {Brownstein}, {Comparat}, {Kneib}, {Malanushenko},
  {Malanushenko}, {Oravetz}, {Pan}, {Parejko}, {Schneider}, {Shelden},
  {Simmons}, {Snedden}, {Tanaka}, {Weaver}, \& {Yan}}]{Thomas.etal.2013a}
{Thomas}, D., {Steele}, O., {Maraston}, C., {et~al.} 2013, \mnras, 431, 1383

\bibitem[{{Tremaine} {et~al.}(2002){Tremaine}, {Gebhardt}, {Bender}, {Bower},
  {Dressler}, {Faber}, {Filippenko}, {Green}, {Grillmair}, {Ho}, {Kormendy},
  {Lauer}, {Magorrian}, {Pinkney}, \& {Richstone}}]{Tremaine.etal.2002a}
{Tremaine}, S., {Gebhardt}, K., {Bender}, R., {et~al.} 2002, \apj, 574, 740

\bibitem[{{Volonteri} {et~al.}(2013){Volonteri}, {Sikora}, {Lasota}, \&
  {Merloni}}]{Volonteri.etal.2013a}
{Volonteri}, M., {Sikora}, M., {Lasota}, J.-P., \& {Merloni}, A. 2013, \apj,
  775, 94

\bibitem[{{Wall} \& {Jenkins}(2003)}]{Wall.Jenkins.2003a}
{Wall}, J.~V., \& {Jenkins}, C.~R. 2003, {Practical Statistics for
  Astronomers}, ed. R.~{Ellis}, J.~{Huchra}, S.~{Kahn}, G.~{Rieke}, \& P.~B.
  {Stetson}

\bibitem[{{Wild} {et~al.}(2010){Wild}, {Heckman}, \&
  {Charlot}}]{Wild.Heckman.Charlot.2010a}
{Wild}, V., {Heckman}, T., \& {Charlot}, S. 2010, \mnras, 405, 933

\bibitem[{{Zakamska} {et~al.}(2016){Zakamska}, {Lampayan}, {Petric}, {Dicken},
  {Greene}, {Heckman}, {Hickox}, {Ho}, {Krolik}, {Nesvadba}, {Strauss},
  {Geach}, {Oguri}, \& {Strateva}}]{Zakamska.etal.2016a}
{Zakamska}, N.~L., {Lampayan}, K., {Petric}, A., {et~al.} 2016, \mnras, 455,
  4191

\end{thebibliography}

\label{lastpage}

\end{document}